\documentstyle[12pt]{article}
\begin{document}
\title{Four-quark state in QCD}
\author {{\small Ailin Zhang\thanks{Email address: zhangal@itp.ac.cn}}\\
{\small CCAST (World Laboratory), P. O. Box 8730, Beijing, 100080}\\ 
{\small Institute of Theory Physics, P. O. Box 2735, Beijing, 100080, 
P. R. China}\\}
\date {}
\maketitle
\begin{center}
\begin{abstract}
The spectra of some $0^{++}$ four-quark states, which are composed of
${\bar qq}$ pairs, are calculated in QCD. The light four-quark
states are calculated using the traditional sum rules while
four-quark states containing one heavy quark are computed in HQET. For
constructing the interpolating currents, different couplings of the
color and spin inside the ${\bar qq}$ pair are taken into account. It
is found that the spin and color combination has little effect on the
mass of the four-quark states.
\end{abstract}
\end{center}
\section{Introduction}
\indent
\par The introduction of color and the development of QCD explain the
classification and many spectroscopic characteristics of hadrons, they
predict the possibility of the existence of exotic hardonic structures
such
as glueball and multiquark states too. However, in contrast
with the normal hardons, the existence of multiquark states is neither
forbidden nor required by color confinement, so the identification of
multiquark states is an interesting topic. It is possible too that
four-quark configurations lead to hadron-hadron potentials\cite{barn},
which play an important role in final state interactions. Therefore,
even though there exists no multiquark state, it is helpful to study
multiquark configurations to understand final state 
interactions\cite{godfrey}.
\par The search for exotics has gone on for a long time, especially
in light hadrons energy region. As we know, there are several ambiguous
resonances among the light resonances, such as the $\sigma(400-1200)$,
$a_0(980)$, $f_0(980)$, $f_0(1500)$, $f_J(1710)$\cite{epj}, etc.
There are many analyses of these resonances. However, since we know
little about glueball and multiquark states, especially because they
are complicated by mixing among these hadron states, we could not identify
them at
present. These light hadron states need further investigation. Not until
we have complete knowledge about the normal hadron, glueball and
multiquark state, can we classify the hadron zoo unambiguously.
Experimentally, there are several four-quark candidates\cite{epj}, such as
$a_0(980)$, $f_0(980)$, $f_1(1420)$, $f_2(1565)$, and $\Psi(4040)$ 
etc. The $a_0(980)$ and $f_0(980)$ lie below the threshold of $K\bar
K$; the rest lie below the threshold of some other meson pairs($K\bar
K^*$, $\omega\omega$, and $D^*\bar D^*$) too. In practice, their
characteristics of decay imply intensely their multiquark bound state
role.   
\par Theoretically, the four-quark system has previously been studied in
the frameworks of the bag model\cite{bag}, of nonrelativistic potential
models\cite{barn},\cite{potential}, and from many other points of
view\cite{other}, but the conclusions for the existence of four-quark
state are quite model dependent, and few investigations are based on QCD.
The
$a_0(980)$ and $f_0(980)$ have been interpreted as the four-quark states
in many papers\cite{bag},\cite{four}, but none of them has been
definitely confirmed as the four-quark state by experiments so far. It is
widely believed that the QCD sum rule is a capable nonperturbative method
to extract the properties of hadrons, but there exist few works on
multiquark states\cite{sum}, especially systematic calculations of
four-quark states. So we hope to examine the properties of four-quark
states with this method in this paper. 
\par Since there are four quarks(including antiquarks) in this
multibody system, the analysis is much more complicated than that for the
normal hadron. Apart from the complication of dynamical interaction among
the quarks, the couplings of color, spin, and flavor among the quarks are
not unique either. It is obvious that the four-quark state can consist of
diquark-antidiquark or $q{\bar q}$ pairs(meson-meson like). We will
concentrate our analysis on the simple $0^{++}$ meson-meson-like
four-quark states only.
\par The spin of the quarks in each $q{\bar q}$ pair can couple to both
singlets and triplets, while color can couple to both singlets and
octets, so there are different combinations for the spin and color between 
quarks. The QCD sum rule is based on the assumption of the existence of
bound
state and resonance, but it cannot distinguish the interaction between
quarks inside the hadron directly. However, we try to construct a
different
sum rule with different current, in which couplings of the spin and
color are different, to detect some information about the four-quark
states. For light quark systems, it may be insufficient to predict the
characteristics of the interaction inside the hadron. However, in the case
of
heavy quark systems, it is possible to study the interactions using the 
sum rule from the analysis of the wave function of this system. For the
presence of interactions, different four-quark states with the same
overall quantum numbers will mix with each other, which makes our analyses
much more difficult. Accordingly, the mixing between these currents
hasnot been taken into account either. To find out any difference
between the light four-quark states and heavy four-quark states
explicitly, we give the sum rules and conclusions for them, respectively.
Surely, we have no intention of identifying the $a_0(980)$ and $f_0(980)$
only through analyses of the spectra of four-quark states, and no
attention has been placed on them. 
\par This paper is organized as follows. The features of four-quark state
have been analyzed in Sec. 2. In Sec. 3, the sum rules for the
$0^{++}$ light four-quark states have been constructed. The heavy
four-quark states with one heavy quark have been analyzed in heavy quark 
effective theory(HQET) in Sec. 4. We
give the numerical results of the spectra in Sec. 5. The last section is
reserved for the conclusion and discussion. 
\section{Review of the features of four-quark state}
\indent
\par In the bag model\cite{bag}, once the confinement has been imposed
on by the boundary condition and four quarks inside the bound state
have been arranged symmetrically about  their center of mass, the spectra
and dominant decay
couplings were calculated. It was found that $q^2{\bar q^2}$
resonances were generally too broad and heavy to show up as bumps in mass
spectra except the $0^{++}$ state, which can be seen in phase shift
analyses. The lighter $0^{++}$ four-quark states were predicted to couple
strongly to two pseudoscalars while the heavier coupled strongly to two
vectors. The observed $a_0(980)$ and $f_0(980)$ have been identified as
the isospin-1 and isospin-0 four-quark states, respectively.
\par In the nonrelativistic potential model\cite{barn},\cite{potential},
after the introduction of the color dependent confinement force and
hyperfine
interactions, the existence of four-quark state is predicted as a
dynamical solution of the Schr$\ddot{o}$dinger equation. In contrast
with the prediction of bag model, it was found that normally the ground
state of this four-quark system consisted of two free mesons except for
the $K{\bar K}$ system(named for the $K{\bar K}$ molecule), which was 
in fact a
weakly bound $0^{++}$ state. There does not exist a rich discrete
spectrum of four-quark states either. The $a_0(980)$ and $f_0(980)$ have
been interpreted as this kind of isospin-1 and isospin-0 $K{\bar K}$
molecule, respectively. There are some other points of view
too\cite{other}.
\par For the  combination of four color quarks into a color singlet
hadron, there exist different ways. We can combine two color triplet q's
into a color $6$ or ${\bar 3}$; similarly, we can combine two antitriplet
${\bar q}$'s into $3$ or ${\bar 6}$. Therefore, there are two ways to
combine the four quarks $qq{\bar qq}$ into the final color singlet:
$3{\bar 3}$ and $6{\bar 6}$. The four-quark states can consist of two
$q{\bar q}$ pairs too; then we can combine the $q{\bar q}$ pair into color
singlet $1$ and color octet $8$, and obtain the final color singlet for
the four-quark states from color combination: $11$ and $88$. It is more
complicated because the $3{\bar 3}$ $6{\bar 6}$ couplings can mix to give
the
$11$-$88$ color configurations.
\par Neither $\vec{L}$ nor $\vec{S}$ is conserved in a relativistic
quark model. Nevertheless, if we consider only the S-wave sector, the
algebra generated by the states and their currents is an $SU(2)$.
Therefore, the spin couplings between two quarks can be symmetric
triplet and asymmetric singlet. For example, the four-quark scalar
states can be considered either as bound states of two ordinary
pseudoscalar mesons or as bound states of two ordinary vector mesons.
As for the flavor combination, the literature\cite{bag} has a detailed
description. The special character of the four-quark states may lie in the
existence of flavor exotics, such as $Q=2$ or $S=2$, which may be our
best chance to find genuine multiquark states.  
\par It was suggested too that, similar to the bound states below the
$K{\bar K}$ threshold, bound states $D{\bar K}$ and $DK$ should exist near
and possibly below the $DK$ threshold\cite{other}. However, except for
the shortness of experimence, there existed few quantitative calculations
of this kind of heavy quark system either. The development of heavy
quark effective theory provides us a capable tool to deal with with these
systems.
\par In the analyses of the literatures mentioned above, all the 
interactions were put into theory by hand, so conclusions about the
properties of interactions between quarks inside four-quark states are
not conclusive either. Taking into account the complication of
couplings of spin and color inside the possible physical four-quark
states, we try to construct sum rules with different currents to explore
them. No matter how much we can extract from these analyses, they are
helpful to us from the point of view of the sum rule itself. For flavor,
all
the following calculations are kept in the unbroken symmetric $SU(3)$. 
\par 
\section{Sum rules for the light four-quark state}
\indent
\par First, let us consider the light $q{\bar q}$ pairs four-quark
states, in which the color couples to a singlet. For the scalar bound
states,
they can be regarded as either two bound pseudoscalars or two two bound
vectors. So we choose the current as
\begin{eqnarray}\label{j1}
j_1(x)=(\bar q\Gamma\Lambda^m q)(\bar q\Gamma\Lambda^n q)(x),
\end{eqnarray}
where for the pseudoscalar quark pairs, $\Gamma=\gamma_5$, while
$\Gamma=\gamma_\mu$ for the vector pairs. $\Lambda^m$ is the generator of
flavor $SU(3)$. 
\par To obtain the operator product expansion, two kinds of Feynman
diagram should be taken into account: one is unbound while the other is
bound. For the pseudoscalar current, the contribution of the bound one is
only $1/12$ of the unbound one; it is the same case for other currents(the
suppressed factor may not be $1/12$) too. Therefore, we will omit the
contributions of bound diagram in the following calculations. All our 
calculations are in the x representation\cite{shifman}, and only those
terms contributing to the sum rules after Borel transformation are kept in
the following formulas.
\par The operator product expansion for the correlation function with
pseudoscalar pairs inside the currents can be expressed as
\begin{eqnarray}
\Pi(q^2)&=&i\int dx e^{iqx}\langle
T\{j_1(x),j^\dagger_1(0)\}\rangle,\\\nonumber
&=&-A(q^2)^4\ln (-q^2)-B(q^2)^2\ln (-q^2)-Cq^2\ln (-q^2) ,
\end{eqnarray}
where A, B and C correspond to the perturbative contribution, two-gluon
condensate, and four-quark condensate, respectively, while two-quark
condensate vanishes.
\par For $\Gamma=\gamma_5$, 
\begin{eqnarray}
A={1\over 163840\pi^6}, B={3\langle\alpha G^2\rangle \over 2^{11}\pi}, 
C={\langle\bar qq\rangle^2 \over 64\pi^2},
\end{eqnarray}
while 
\begin{eqnarray}
A={1\over 40960\pi^6}, B={\langle\bar qq\rangle^2 \over 32\pi^2}
\end{eqnarray}
in the case of $\Gamma=\gamma_\mu$, where the two-gluon condensate
vanishes
also.
\par The imaginary part of the correlation functions can be represented as
\begin{eqnarray}
Im\Pi(s)=\pi f^2_0(m^4)^2\delta(s-m^2)+\pi(As^4+Bs^2+Cs)\theta(s-s_0),
\end{eqnarray}
where the first term is from the lowest lying bound state or resonance
and the second one is from higher resonances or continuum states.
\par Similarly, the $q{\bar q}$ pair can couple to a color octet too, and
the currents are chosen as
\begin{eqnarray}
j_2(x)=f^{ab_1c_1}f^{ab_2c_2}(\bar q^{b_1}\Gamma\Lambda^m q^{c_1})
(\bar q^{b_2}\Gamma\Lambda^n q^{c_2})(x),
\end{eqnarray}
where $\Gamma$ is the same as the definition below formula (\ref{j1}).
\par The coefficients of the correlation functions for $\Gamma=\gamma_5$
and $\Gamma=\gamma_\mu$ are
\begin{eqnarray}
A={3\over 163840\pi^6}, B={9\langle\alpha G^2\rangle \over 2^{11}\pi},
C={3\langle\bar qq\rangle^2 \over 64\pi^2},
\end{eqnarray}
and 
\begin{eqnarray}
A={3\over 40960\pi^6}, C={3\langle\bar qq\rangle^2 \over 8\pi^2},
\end{eqnarray}
respectively.
\par Then after equating the quark sides with the hadron sides with
the dispersion relation, we obtain the mass of the four-quark state,
\begin{eqnarray}
m^2(s0,\tau)={R_{k+1}(s_0,\tau)\over R_k(s_0,\tau)},
\end{eqnarray} 
where $s_0$ is the continuum threshold, $\tau$ is the Borel transformation
variable and\cite{hjz} 
\begin{eqnarray}
R_k(\tau,
s_0)&=&\frac{1}{\tau}\hat{L}[(q^2)^k\{\Pi(Q^2)-\sum\limits_{k=0}^{n-1}a_k(q^2)^k\}]
-\frac{1}{\pi}\int_{s_0}^{+\infty}s^k
e^{-s\tau}Im\Pi^{\{pert\}}(s)ds\\\nonumber
&=&\frac{1}{\pi}\int_{0}^{s_0}s^k e^{-s\tau}Im\Pi(s)d s.
\end{eqnarray}
\section{Sum rules for the heavy four-quark state}
\indent
\par In this section, we will discuss the scalar four-quark systems with
one heavy quark. With the same consideration as the previous section, the
interpolating current corresponding to the color singlet of a
quark-antiquark
pair is chosen as
\begin{eqnarray}
j_3(x)=(\bar q\Gamma h_v)(\bar q\Gamma\Lambda^m q)
\end{eqnarray}
where $q(x)$ is the light quark field, $h_{\it v}(x)$ is the heavy quark
effective field, and $v$ is the velocity of the heavy quark. 
\par Then, we construct the correlation function as 
\begin{eqnarray}
\Pi(\omega)
&=&i\int d^4x e^{iqx} \langle 0| T\{j_3(x),j^\dagger_3(0)\}|0 \rangle,
\end{eqnarray}
where
\begin{eqnarray}
\omega=2q\cdot v.
\end{eqnarray}
\par After twice suitable Borel transformations, the imaginary part of it
is obtained
\begin{eqnarray}
Im\Pi(\tau)=A\tau^8+B\langle\bar qq\rangle\tau^5+C\langle\alpha_s G^2
\rangle\tau^4+D\langle\bar qq\rangle^2\tau^2. 
\end{eqnarray}
\par When $\Gamma=\gamma_5$,
\begin{eqnarray}
A={9\over 2^9\cdot 8!\pi^5}, B={-3\over 2^7\cdot 5!\pi^3}, C={9\over
2^{10}\cdot 4!\pi^4}, D={1\over 2^6\pi}.
\end{eqnarray}
\par For $\Gamma=\gamma_\mu$,
\begin{eqnarray}
A={9\over 2^7\cdot 8!\pi^5}, B={-3\over 2^6\cdot 5!\pi^3}, D={1\over
2^5\pi},
\end{eqnarray}
and C vanishes. To obtain the results above, we have taken use of the
infinite heavy quark mass limit.
\par The current with the color octet quark-antiquark pair is chosen as
\begin{eqnarray}
j_4(x)=f^{ab1c1}f^{ab2c2}(\bar q^{b1}\Gamma h^{c1}_v)(\bar
q^{b2}\Gamma\Lambda^m
q^{c2}).
\end{eqnarray}
\par The coefficients of $A$, $B$, $C$ and $D$ for
$\Gamma=\gamma_5,\gamma_\mu$ are
\begin{eqnarray}
A={27\over 2^9\cdot 8!\pi^5}, B={-9\over 2^7\cdot 5!\pi^3}, C={27\over
2^{10}\cdot 4!\pi^4}, D={3\over 2^6\pi},
\end{eqnarray}
and
\begin{eqnarray}
A={27\over 2^7\cdot 8!\pi^5}, B={-9\over 2^6\cdot 5!\pi^3}, D={3\over 
2^5\pi},
\end{eqnarray}
respectively.
\par On the phenomenal side, the correlation function is expressed as 
\begin{eqnarray}
\Pi(\omega) = {F^2_{H^+} \over (2\Lambda-\omega)}
+\int_{\omega_c}^{\infty} d\omega'
{Im\Pi_s(\omega') \over \omega'-\omega} ,
\end{eqnarray}
where the first term on the right side is the dominant pole term resulting
from the lowest lying resonance contribution, the second term represents
the
contribution of the continuum state and higher resonances, and $\omega_c$
is the
continuum threshold. 
\par Making use of the dispersion relations for the correlation functions 
to equate the quark sides with hadron sides, we obtain
\begin{eqnarray}\label{spect}
{F^2_{H^+}\over (2\Lambda-\omega)} 
= &{1\over\pi}\int_{0}^{\omega_c} d\omega' 
{Im\Pi(\omega') \over \omega'-\omega} ;
\end{eqnarray}
after Borel transformation, they are turned into 
\begin{eqnarray}
F^2_{H^\pm} e^{-2\Lambda/T}
= &{1\over\pi}\int_{0}^{\omega_c} d\omega'
Im\Pi(\omega') e^{-\omega'/T} ,
\end{eqnarray}
where $T$ is the Borel transformation variable. 
So the $\Lambda$ can be determined as
\begin{eqnarray}\label{lam}
2\Lambda = {\int_{0}^{\omega_c}d\omega'
\omega'Im\Pi(\omega')e^{-\omega'/T}\over \int_{0}^{\omega_c}d\omega'
Im\Pi(\omega')e^{-\omega'/T}}
\end{eqnarray}
\section{Numerical results of the spectra for the four-quark states}
\indent
\par In this section, we will give the numerical results of the spectra 
for the four-quark states. To proceed with the calculation, the mass of
the b
and c quarks are chosen as $4.7$ GeV and $1.3$ GeV, respectively. The
condensates are chosen as
\begin{eqnarray} 
\langle 0|\bar qq|0 \rangle = -(0.24 GeV)^3,
\langle 0|\alpha_s G^2|0 \rangle = 0.06 GeV^4.
\end{eqnarray}
\par A few words should be given about the technical details first. In
the light quark case, we tried the calculation with $s_0=1.0, 1.5$
and $2.5$ GeV, respectively. The mass square of them is displayed in
Figs. 1 and 2 though the platform is not satisfactory. For the heavy
four-quark states, the $2\Lambda$ obtained by us is shown in Figs. 3 and
4,
where the $\omega_c$ are chosen as $2.0$, $3.0$, and $4.0$ GeV,
respectively. To find which $s_0$ or $\omega_c$ is the suitable one for
our sum rules, the ordinary criteria\cite{hjz} of the determination of the
threshold and platform are taken into account. All the results are
collected in Table I. 
\ref{table1}. 
\begin{table}[b]
\caption{Mass of the scalar four-quark states from different spin and
color combination.}
\label{table1}
\begin{tabular}{c c c c c c}
(GeV)&$\gamma_5\bigotimes singlet(c)$  &  $\gamma_\mu\bigotimes
singlet(c)$ &
$\gamma_5\bigotimes octet(c)$ & $\gamma_\mu\bigotimes octet(c)$
\\
\hline
 light($m$)        & 0.74-0.87 & 0.71-0.89 & 0.74-0.90 & 0.71-0.87 &\\
 heavy($2\Lambda$) & 0.95-1.26 & 0.97-1.24 & 0.95-1.24 & 0.97-1.24 &\\
\end{tabular}
\end{table}
\par From the results obtained here, the light four-quark states are found
to be light. When the mass of the s quark is taken into account, the
conclusion will
not change. So it is reasonable to search for four-quark states in the
light hadron regions, while the heavy four-quark states with one $c$ or
$b$
quark are found to lie below $2.0$ GeV or $5.5$ GeV, respectively. The
mass prediction\cite{other} of the $D{\bar K}$ or $DK$ four-quark
states is higher than the result here.
\par As for the effect of interactions between quarks on the mass of
four-quark states, in principle, it cannot be studied through sum rule
methods directly. There is no correspondence between the relative
interpolating currents and the bound states. Nevertheless, we can proceed
with the sum rules process with different currents as we did previously.
Especially, we believe that sum rules may be capable of finding some
information about the interactions in heavy quark systems, where there
exists a nonrelative limit. It is found that the couplings of spin and
color inside the $q{\bar q}$ pair have little effect on the
mass determination of them. The four different combinations between the
spin and color inside the currents result in similar masses in both
light and heavy quark systems.
\par Through the lessons from analyses of mesons and glueballs, we
have learned that perturbative contributions play a dominant role in
the sum rules. However, in the present case, the four-quark condensate
plays the dominant role, and all the other contributions(including the
perturbative term) are negligible. The results obtained here infer
that either the interactions inside the four-quark states are some what
special or it is not suitable to deal with these states using QCD sum
rules. 
\par For systems with one heavy quark, there exists some symmetries
and a nonrelativistic model may work well. Therefore, sum rule
analyses with different combinations of spin and color in the
interpolating currents in heavy quark effective theory may explore some
physical information about the four-quark states. It can be seen that the
platform for these systems is much better than that for light quark
systems. Especially, in contrast with the light quark case, the
perturbative contribution dominates the sum rule here, which means that
application of the HQET sum rule to heavy four-quark states is suitable.
\section{Conclusion and discussion}
\indent
\par The spectra of $(q{\bar q})(q{\bar q})$ meson-meson like four-quark
states have been calculated using QCD sum rules. Four-quark states
with light quarks are found below $1.0$ GeV. The mass of heavy four-quark
states with a c or b quark has been predicted too. 
 \par In other models, to extract the properties of hadrons, the
characteristics of interactions were put into the theory by hand.
Unlikely,
both the hadron properties and the interactions inside are predicted on 
basis of QCD in the framework of sum rules. For four-quark
states, whether they are compact states such as the ordinary mesons or
weakly
bound states such as molecules is of the key concern with us. Moreover,
the
complicated couplings of spin and color between the quarks inside the
bound states need not be speculated about either. A possible choice is
to construct suitable currents corresponding to the topic with which we 
are concerned,
and the analysis of the reasonableness for choosing currents is important
to
us. 
\par In this paper, we proceeded with sum rule analyses with different
currents, where different combinations of spin and color
inside the $q{\bar q}$ pair have been tried. In fact, it may be a good way
to detect the physical couplings of spin and color inside the bound states
in heavy quark systems. Even though there is no correspondence between
the currents and physical bound states at all, it is necessary to find
which interpolating current is the right one for the special question.
Previous results confirm that different combinations of spin and color
inside the currents have little effect on the mass determination of
four-quark states.
\par Experimentally, we cannot confirm the four-quark state yet; we
have observed only some final strong coupling meson-meson channel. For
this reason, we consider only the $(q{\bar q})(q\bar q)$
four-quark currents in our sum rules. For a complete sum
rule analysis, the $(qq)(\bar q \bar q)$ four-quark states should be
taken into account too. Moreover, besides the mixing between ordinary
hadrons and four-quark states, there exists mixing between the four-quark
states with the same overall quantum number also. Accordingly, the mixing
between different currents should be considered. Unfortunately, we know
little about the mixing, both in QCD and in some models. This
problem has remained beyond the scope of this paper.\\

\vspace{1.0cm}
{\bf Acknowledgment}: 
The author is grateful to Professor Yuan Ben Dai for useful discussions.

\newpage
\par
{\huge\bf Figure caption}\\
\par
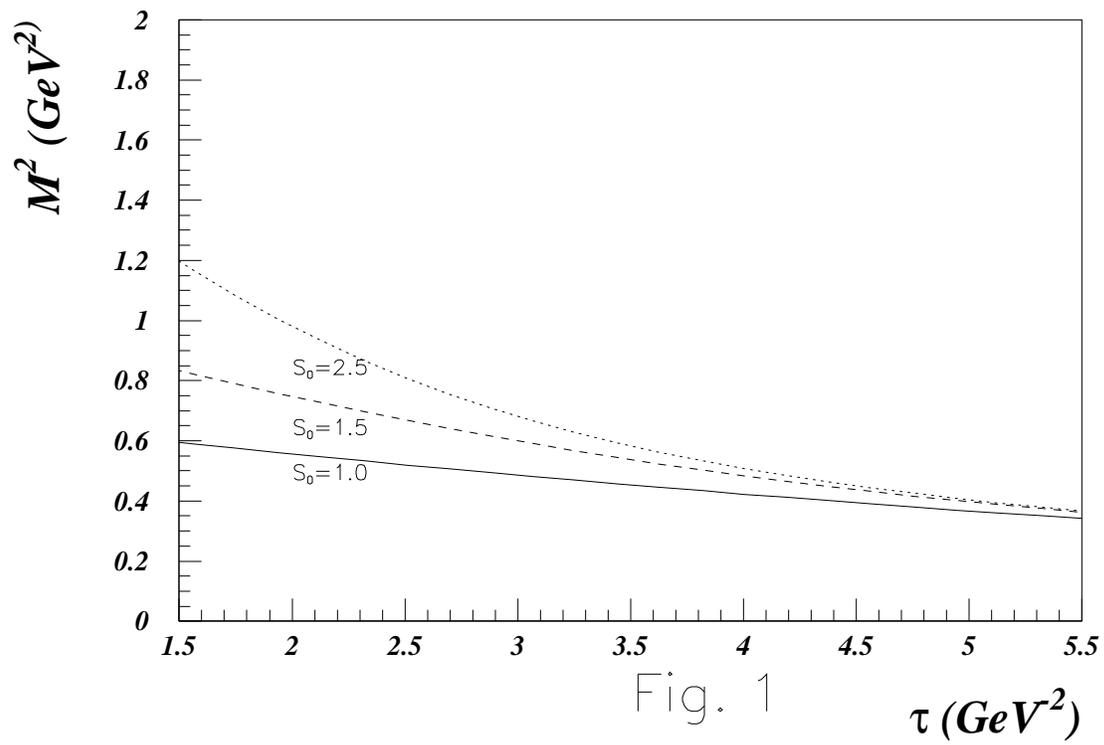
Fig. 1: Mass square of $0^{++}$ four-quark state from current $j_1(x)$
versus Borel variable $\tau$.\\
\par
\par
Fig. 2: Mass square of $0^{++}$ four-quark state from current
$j_2(x)$ versus Borel variable $\tau$.
\par
\par
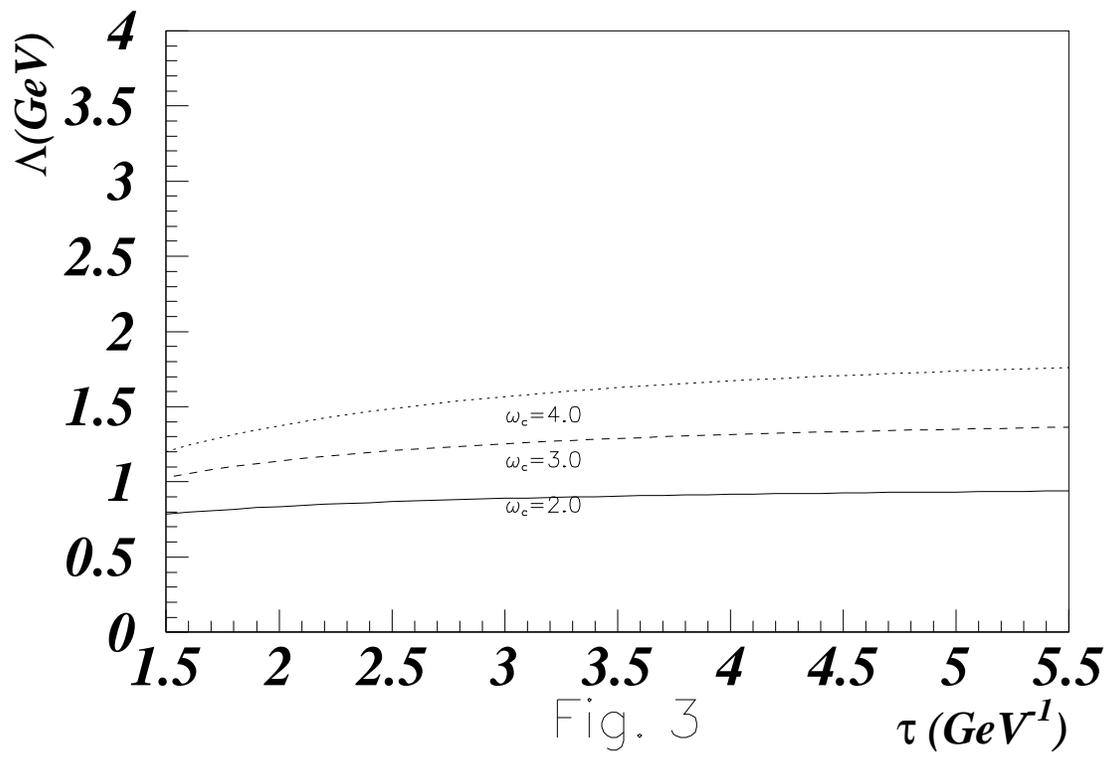
Fig. 3: $2\Lambda$ of  $0^{++}$ heavy-light four-quark state from 
current $j_3(x)$ versus Borel variable T.\\
\par
\par
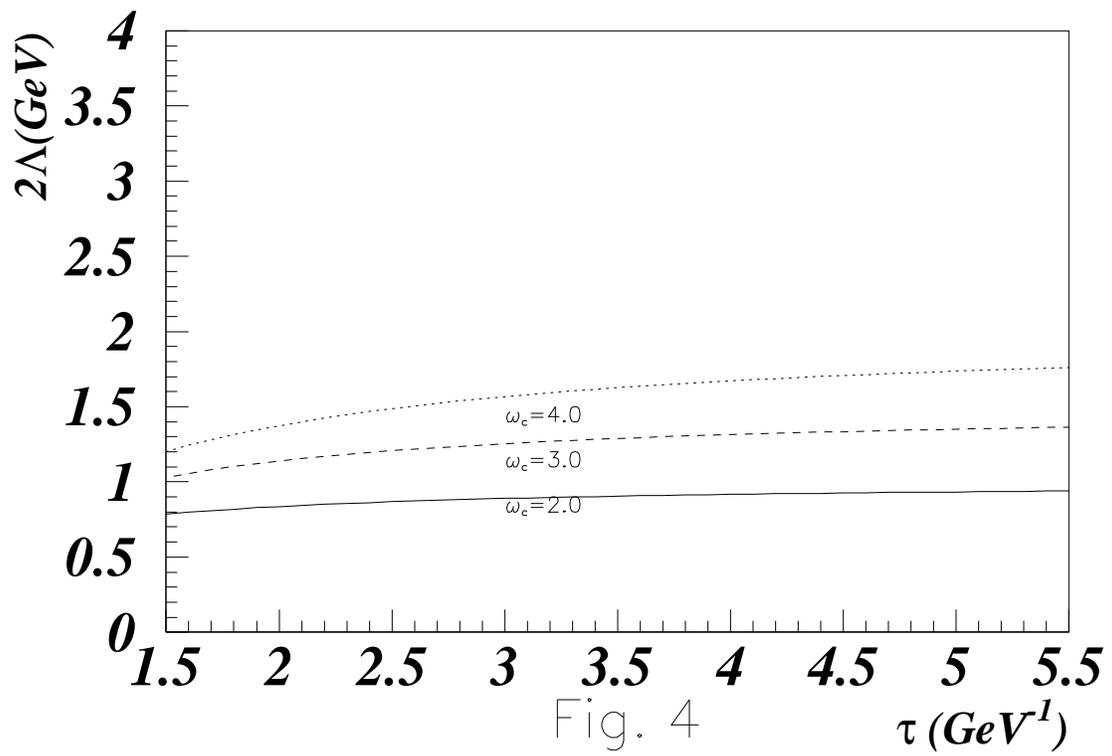
Fig. 4: $2\Lambda$ of $0^{++}$ heavy-light four-quark state from 
current $j_4(x)$ versus Borel variable T.
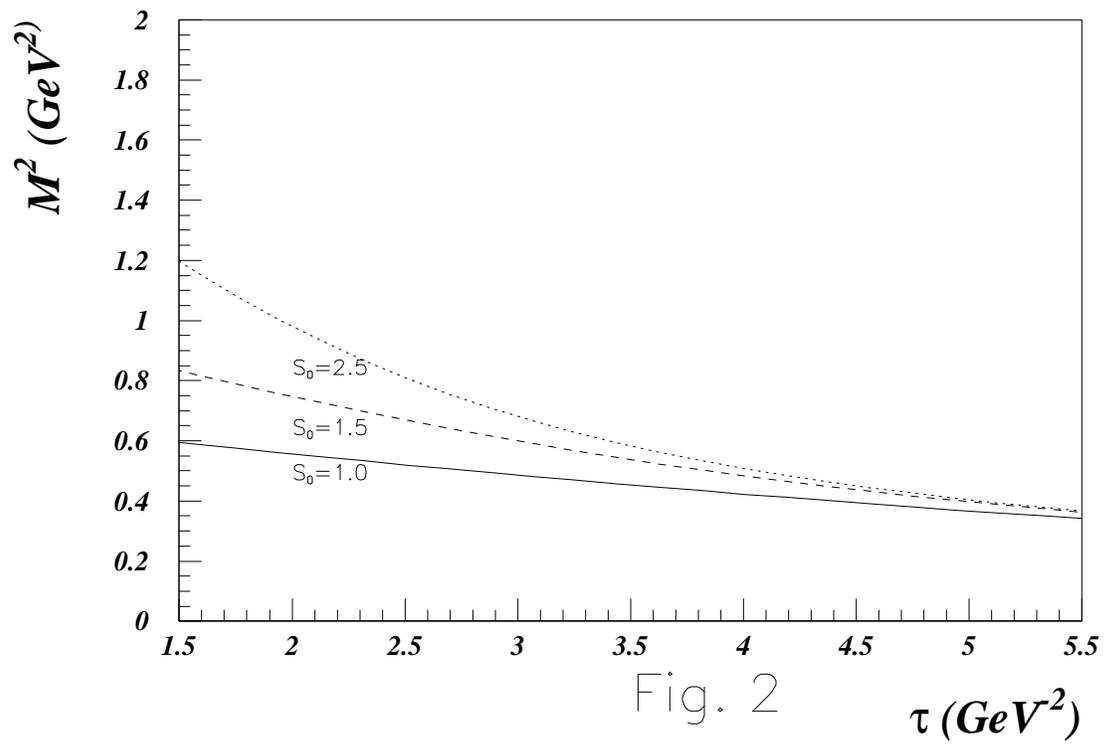
\end{document}